\documentclass{PoS}

\title{4C~02.27: what is inside a double-double radio quasar?}

\ShortTitle{4C~02.27: what is inside a double-double radio quasar?}

\author{\speaker{S\'andor Frey}%
         \thanks{The EVN is a joint facility of European, Chinese, South African, and other radio astronomy institutes funded by their national research councils. The e-VLBI research infrastructure in Europe is supported by the European Community's Seventh Framework Programme (FP7/2007-2013) under grant agreement no.\ RI-261525 NEXPReS. This work was supported by the European Community's Seventh Framework Programme, Advanced Radio Astronomy in Europe, grant agreement no.\ 227290, and the Hungarian Scientific Research Fund (OTKA, grant no.\ K72515). This research has made use of the NASA/IPAC Extragalactic Database (NED) which is operated by the Jet Propulsion Laboratory, California Institute of Technology, under contract with the National Aeronautics and Space Administration.}\\
        F\"OMI Satellite Geodetic Observatory, P. O. Box 585, H-1592 Budapest, Hungary\\ 
        MTA Research Group for Physical Geodesy and Geodynamics, Budapest, Hungary\\
        E-mail: \email{frey@sgo.fomi.hu}}

\author{Zsolt Paragi\\
        Joint Institute for VLBI in Europe, Postbus 2, 7990 AA Dwingeloo, The Netherlands\\
        MTA Research Group for Physical Geodesy and Geodynamics, Budapest, Hungary\\
        E-mail: \email{zparagi@jive.nl}}

\abstract{Recently Jamrozy et al. (2009) identified 4C~02.27 (J0935+0204) as the first possible example of a double-double radio source which is optically identified with a quasar (i.e. not a galaxy), at the redshift of $z$=0.649. The overall projected angular size of the radio source reaches about $1.5^{\prime}$, with a prominent ``core'' component in the centre. The two opposite pairs of radio lobes might indicate two periods of episodic activity. We report on our short exploratory 1.6-GHz Very Long Baseline Interferometry (VLBI) observations of the innermost radio structure of the quasar, conducted with the electronic European VLBI Network (e-EVN) on 2009 September 30. These revealed a milliarcsecond-scale compact source which is the base of the approaching one of the two symmetric relativistic jets currently supplying the hot spots in the inner pair of the arcsecond-scale radio lobes in 4C~02.27.}

\FullConference{10th European VLBI Network Symposium and EVN Users
Meeting: VLBI and the new generation of radio arrays \\
		 September 20-24, 2010\\
		 Manchester, UK}

\begin{document}

\section{Introduction}
The activity of galactic nuclei in the radio lasts for up to $\sim$$10^6$ years, therefore is episodic compared to the whole lifetime of an object, which could be orders of magnitude longer. The radio jet activity could also be recurrent, as suggested by several cases where two or even three aligned pairs of radio lobes are observed in the opposite sides of a central active galactic nucleus (AGN), at different separations. The relic lobes provide ``historical'' information on different activity cycles via e.g. their spectra, brightness, polarisation properties, size, and location. A comprehensive review of the observational evidence of recurrent activity in AGNs is found in \cite{Saik09}, where 15 examples of double-double radio galaxies and one triple-double radio galaxy are collected from the literature.  

\begin{figure}[!h]
\centering
\includegraphics[bb=40 167 568 639, height=75mm, angle=0, clip=]{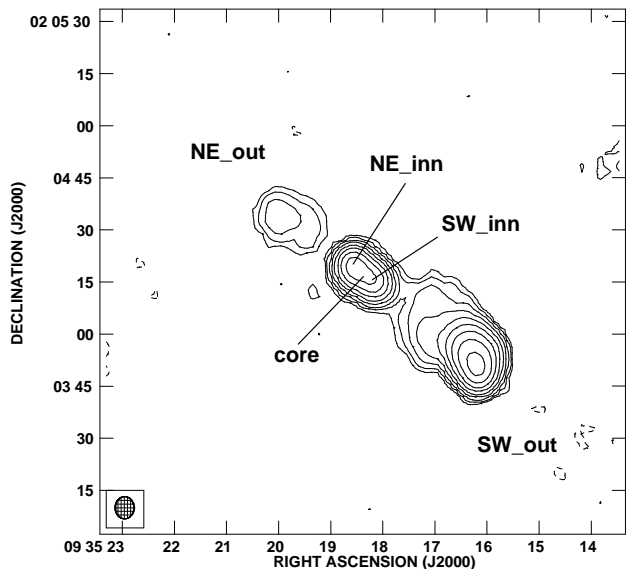}
\caption{1.4-GHz radio image of 4C~02.27 from the VLA FIRST survey \cite{Whit97}. The lowest contours are drawn at $\pm0.56$~mJy/beam (3$\sigma$ image noise). The positive contour levels increase by a factor of 2. The peak brightness is 205.19~mJy/beam. The Gaussian restoring beam ($6.4^{\prime\prime} \times 5.4^{\prime\prime}$, major axis position angle $0^{\rm o}$) is shown in the bottom-left corner. The components are labelled according to \cite{Jamr09}.}
\label{FIRST-image}
\end{figure}

Recently \cite{Jamr09} identified 4C~02.27 (J0935+0204) as the first possible example of a double-double radio {\it quasar} (DDRQ). Its redshift is $z$=0.649 as measured in the Sloan Digital Sky Survey \cite{sdss}. The outer pair of lobes can easily be identified in the 1.4-GHz radio image (Fig.~\ref{FIRST-image}) taken in the US National Radio Astronomy Observatory (NRAO) Very Large Array (VLA) Faint Images of the Radio Sky at Twenty-centimeters (FIRST) survey \cite{Whit97}. The inner lobes, of which the north-east (NE) one is the brightest, are not well resolved in the FIRST image and blended together with a central ``core'' component. A VLA A-array image obtained at the same frequency \cite{Hint83} indicate that the separation of the inner lobes is nearly $10^{\prime\prime}$.  

There is a large brightness and arm-length asymmetry between the inner and outer pairs of lobes. This suggests that environmental asymmetries and relativistic motion together determine the appearance of this source. An alternative explanation could be a jet reorientation between the corresponding two cycles of activity, although it is less likely because of the special geometry required by the good alignment of the inner and outer lobes in the sky \cite{Jamr09}. It was found that the a bulk plasma velocity of $\sim$0.4$c$ and a jet inclination of $\sim$$40^{\rm o}$ with respect to the line of sight are consistent with the observed arcsecond-scale inner structure. In this scenario, the inner NE lobe is approaching to us.      

While the source extends to almost $1.5^{\prime}$, there is a ``core'' component with a flux density of $\sim$90~mJy at 1.4~GHz, and a rather flat radio spectrum (inferred from other VLA data at higher frequencies: \cite{Swar84,Pric93,Boge94}). The position of the radio core nearly coincides with the quasar's optical position. 4C~02.27 has also been detected in X-rays with ROSAT \cite{Voge00}.

\section{Observations, results and discussion}

The source has never been imaged with Very Long Baseline Interferometry (VLBI) before. With the experiment described below, our goal was to detect any possible compact radio feature in the central component on $\sim$10-milli-arcsecond (mas) angular scale with the European VLBI Network (EVN). To this end, short 2-h exploratory e-EVN observations were initiated. The project RSF03 took place on 2009 September 30, involving radio telescopes at Effelsberg (Germany), Medicina (Italy), Onsala (Sweden), Toru\'n (Poland), Jodrell Bank Mk2, Cambridge, Knockin (United Kingdom), and the phased array of the Westerbork Synthesis Radio Telescope (The Netherlands). The observations were done at 1.6~GHz frequency, with the maximum bandwidth of 64~MHz in both left and right circular polarizations. The EVN Data Processor (Joint Institute for VLBI in Europe, Dwingeloo, The Netherlands) performed the correlation in real time.

\begin{figure}[!h]
\centering
\includegraphics[bb=67 176 508 619, height=75mm, angle=270, clip=]{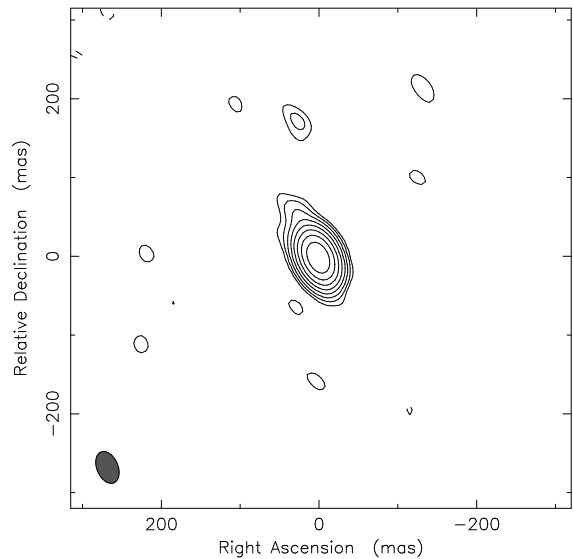}
\caption{1.6-GHz phase-referenced e-EVN image of 4C~02.27. The lowest contours are drawn at $\pm0.25$~mJy/beam. The positive contour levels increase by a factor of 2. The peak brightness is 60.7~mJy/beam. The Gaussian restoring beam (42.5~mas$\times$26.7~mas, major axis position angle $23^{\rm o}$) is shown in the bottom-left corner. The absolute equatorial coordinates of the brightness peak (J2000 right ascension $09^{\rm h}35^{\rm m}18.1946^{\rm s}$ and declination $02^{\rm o}04^{\prime}15.558^{\prime\prime}$) are within the accuracy of 2~mas.}
\label{VLBI-image}
\end{figure}

The target source 4C~02.27 was observed in phase-reference mode, using the nearby calibrator J0930+0034 (1.85$^{\rm o}$ far away). The NRAO Astronomical Image Processing System (AIPS) was used for data calibration and fringe-fitting in a standard way. It turned out that 4C~02.27, a compact source on VLBI scale with $\sim$60~mJy correlated flux density, could itself be fringe-fitted as well. The imaging, self-calibration, and model-fitting were done with the Caltech Difmap package. Here we present the image obtained with phase-referencing (Fig.~\ref{VLBI-image}), which is practically identical to that resulted from direct fringe-fitting to the target source data.  

According to our e-EVN image, the core of 4C~02.27 is compact, slightly extended to NE. Its brightness distribution is well described by a central circular Gaussian component (60.8~mJy flux density, 1.6~mas diameter, full width at half maximum) and  a weak extension (1.6 mJy, 20~mas) at the position angle of $47^{\rm o}$, separated by 32~mas from the central one. The inferred rest-frame brightness temperature of the compact feature is 1.7$\times$$10^{10}$~K. Assuming the equipartition value of 5$\times$$10^{10}$~K for the intrinsic brightness temperature \cite{Read94}, it is consistent with the $\sim$$40^{\rm o}$ inclination suggested by \cite{Jamr09} and a reasonable jet bulk Lorentz factor $\sim$12. The apparently one-sided jet-like structure indicates that the NE side is the approaching one, just as deduced from the appearance of the inner lobes \cite{Jamr09}. 


With VLBI we observe the base of the approaching relativistic jet in 4C~02.27, a currently active radio quasar. The jets provide the energy supply for the hot spots in the {\it inner pair} of the arcsecond-scale radio lobes seen in the VLA images. The asymmetry of the inner lobes is most likely caused by relativistic beaming effects. There is no need to invoke extreme misalignment between the mas-scale and arcsecond-scale structures. At the resolution provided by the e-EVN at 1.6~GHz, we found no sign of an innermost double structure (unlike e.g. \cite{Mare09}, in the case of the radio galaxy B0818+214) which would have indicated a third and most recent cycle of activity in this unique radio-loud quasar.

The appearance of the {\it outer lobes} is probably determined by environmental asymmetries and/or light travel time effects. We see a later evolutionary stage of the fading outer lobe in the approaching side (NE\_out), it could therefore appear fainter than its counterpart on the receding side (SW\_out). Considering an aligned source geometry with $\sim$$40^{\rm o}$ inclination to the line of sight, the light travel time difference is nearly $2\times10^6$~years between the two outer lobes.  

\end{document}